\documentclass[a4paper,english]{article}

\usepackage{amsmath}

\usepackage{bibentry}
\nobibliography{literatur}

\usepackage{nopageno}
\usepackage{latexsym}

\usepackage{graphicx}

\usepackage[closeFloats]{fltpage}

\usepackage{afterpage}
\usepackage{nonfloat}
\usepackage{xcolor}

\usepackage[standard]{ntheorem}
\theoremclass{Definition}
\theoremstyle{break}
\theoremheaderfont{\bf\Large}
\theoremprework{\pagebreak\noindent\hrulefill}
\theorempostwork{\noindent\hrulefill}
\usepackage{hyperref}
\usepackage{xspace}
\usepackage{url}
\usepackage{amsfonts}
\usepackage{amssymb}
\usepackage{amsmath,bm}

\usepackage{listings}
\title{\textsf{Collaborative Software Development on the Web}}

\author{Martin Nordio, H.-Christian Estler, Carlo A. Furia, and Bertrand Meyer \\
	ETH Zurich, Switzerland \\
   \url{firstname.lastname@inf.ethz.ch}
}

\date{\today}

\newcommand{\DOSE}{\textsc{dose}\xspace}

\begin{document}

\maketitle

\begin{abstract}
  Software development environments (IDEs) have not followed the IT
  industry's inexorable trend towards distribution. They do too little
  to address the problems raised by today's increasingly distributed
  projects; neither do they facilitate collaborative and interactive
  development practices. A consequence is the continued reliance of
  today's IDEs on paradigms such as traditional configuration
  management, which were developed for earlier modes of operation and
  hamper collaborative projects. This contribution describes a new
  paradigm: cloud-based development, which caters to the specific
  needs of distributed and collaborative projects. The CloudStudio IDE
  embodies this paradigm by enabling developers to work on a shared
  project repository. Configuration management becomes unobtrusive; it
  replaces the explicit update-modify-commit cycle by interactive
  editing and real-time conflict tracking and management. A case
  study involving three teams of pairs demonstrates the usability of
  CloudStudio and its advantages for collaborative software
  development over traditional configuration management practices.

\end{abstract}


%

\section{Introduction}\label{intro}


The Integrated Development Environment is the software developer's
central tool. IDEs have undergone considerable advances; their
fundamental structure and mode of operation are still, however, what
they were decades ago. In particular, while Internet development has
benefitted from IDEs, the IDE has not benefitted from the Internet; it
remains an essentially personal tool, requiring every member of a
project to work on a different copy of the software under development
and periodically to undergo a painful process of reconciliation.

CloudStudio, the IDE described in this article, brings software
development to the Internet. In recent years ever more human
activities, from banking to text processing, have been ``moved to the
cloud''. CloudStudio does the same for software engineering by
introducing a new paradigm of software development, where all the
products of a software project are shared in a common web-based
repository.

Moving software development to the cloud is not just a matter of
following general trends, but a response to critical software
engineering needs, which current technology does not meet: supporting
today's distributed developments, which often involve teams spread
over many locations, and iterative development practices such as pair
programming and online code reviews; maintaining compatibility between
software elements developed by different team members;
avoiding potentially catastrophic version incompatibility problems;
drastically simplifying configuration management.

CloudStudio brings flexibility to several new facets of software
development, most importantly configuration management (CM): to
replace the traditional and painful update-modify-commit-reconcile
cycle, CloudStudio tracks changes at every location in real time and
displays only the selected users' changes in the integrated editor.
The compiler and other tools are aware of the current user
preferences, and target the version of the code coinciding with the
current view.  CloudStudio also integrates communication tools (a chat
box and Skype), and includes a fully automated verification component,
including both static (proof) and dynamic (testing) tools (see
Section~\ref{sec:verification}). This array of tightly integrated tools makes
CloudStudio an innovative IDE, which can improve the quality and speed of projects involving distributed
teams, and support highly collaborative development practices.

CloudStudio is an ambitious project for which we have built a
prototype, which readers can try out (see Section~\ref{cloudstudio}).
To demonstrate CloudStudio's potential to facilitate distributed
development, we have conducted a case study where three teams of two
programmers modified and extended existing software projects with
CloudStudio and with traditional CM (i.e., Subversion).  Within the
limits given by its limited extent, the case study substantiates the
claims that CloudStudio can facilitate collaborative development
without interfering with the standard habits of programmers.  While
the initial results from the prototype are exciting, many research
challenges remain. This article describes both the current CloudStudio
framework and the open research challenges that lie ahead.

Section~\ref{distributed} presents the challenges of collaborative and
distributed development. Section~\ref{cloudstudio} is an overview of
CloudStudio from the user perspective.  Section~\ref{configuration}
describes CloudStudio's CM model and awareness system.
Section~\ref{case_study} presents a case study used to evaluate
CloudStudio's potential for collaborative development.
Section~\ref{conclusions} summarizes and discusses future work.




\section{Distributed and Collaborative Development}\label{distributed}

Today's software projects are increasingly multipolar. ``Gone are the
days of one-company, one-site projects; most industry developments
involve teams split over several locations, countries,
cultures''~\cite{NordioJosephMeyerTerekhov10}. Such projects involve
not just developers but many other stakeholders with different
backgrounds and needs, from users and managers to testers and
trainers. Organizationally, they no longer limit themselves to a
single location or even a single company but follow talent wherever it
is, increasingly leading to a distributed mode of development.

Such distributed projects raise a full set of new software engineering
challenges~\cite{Meyer06b,HolmstromConchuirAgerfalkFitzgerald06,HerbslebMoitra01}, which 
the standard approaches do not address
well. Examples of these challenges include requirements and 
interface specification
in the context of distributed development. Many failures have been
reported in outsourced and distributed projects, often due not to lack
of technical expertise, but to difficulties in management and
communication. Distributed projects require new methods and
sophisticated tool support to handle the complex interactions between
the many actors involved.

An orthogonal trend that brings its own challenges is the growth of
methods based on iterative, incremental, and highly collaborative
development, such as agile methods.  These approaches advertise
informal collaboration and continuous direct communication between
team members as solutions to the deficiencies of traditional
structured development processes.  Whether and how intense
collaboration is achievable when programmers do not sit in the same
room are, however, open question; and even for developers working at
the same location, tools specifically designed to
facilitate collaborative development are still largely unseen.

A central issue, often playing a major part in project failures, is
configuration management. CM addresses fundamental needs: making sure
that all project members use the same reference versions of every
software element, avoiding conflicts as they change various parts of
the system, avoiding configuration errors (where version $n$ of module
$A$ uses the wrong version of module $B$), avoiding regression errors
(where a previously corrected bug reappears as a result of bad
information flow), allowing the re-creation of a previous version of
the system or one of its modules.

The initial impetus for CloudStudio was our experience with
distributed software development both in the context of a long-running
industry development, distributed over many sites and led by the last
author, and with a university course which we have taught for several
years with a distributed collaborative project involving student teams
from several
universities~\cite{NordioMitinMeyer2010}. We
found that today's tools are badly lacking in support for such
distributed setups:

\begin{itemize}

\item Communication is a critical issue. Tools such as Skype, WebEx,
  GoogleDocs, and wikis are useful but not meant for software
  development.
\item Configuration management, the key day-to-day practical issue, is
  a major hurdle. While CM is essential in any team effort, the tools,
  based on 30-year-old concepts, are heavy to use (requiring constant
  ``update'' and ``commit'' operations) and poorly adapted to modern
  distributed projects. These operations distract developers from
  their truly important tasks. Between the time a developer checks out
  a component and checks it back in, the project manager and the rest
  of the team have no idea of what is happening to it; if two
  developers modify the same component, conflicts will be detected
  late, when they likely are hard to reconcile. There is always a tendency to
  branch, often leading to a catastrophe down the line at the time of
  merging. (Unlike physicists, software developers have their Big
  Bang at the end.)

\end{itemize}

Our vision is a new paradigm for software development, both addressing
the needs of distributed projects and taking advantage of
distribution. The vision is embodied in an experimental distributed
software development environment, CloudStudio, allowing teams to work
on a common product regardless of their geographical location. Instead
of running on each developer's machine, CloudStudio is hosted on the
Web and works on a \textit{shared project repository}. The result is a
radically new approach providing developers and managers, at any time,
with an accurate and up-to-date picture of the entire project. It also
includes a profound rethinking of the fundamental task of
configuration management, which becomes an unobtrusive automatic technique
for keeping track of changes on the developers' behalf and
reconstructing earlier versions on demand.

Characteristics of CloudStudio include:
	
\begin{itemize}

\item Unobtrusive configuration management: CloudStudio gives each
  developer the appearance of having a private copy of the project,
  but the project is ``in the cloud'', its material shared between all
  project members. 
  There is practically no need for traditional update and commit: CM
  happens in the background as a result of editing actions.
\item Awareness system: CloudStudio keeps track of all the changes
  introduced by the developers, and lets any developer display the
  changes of the other developers. CloudStudio allows for compiling
  and verifying the project including/excluding these changes. Thus, a
  developer's modifications do not block others.

\end{itemize}

\section{A Session with CloudStudio}\label{cloudstudio}

\begin{figure}[!t]%
\begin{center}
	\includegraphics[width=.90\textwidth]{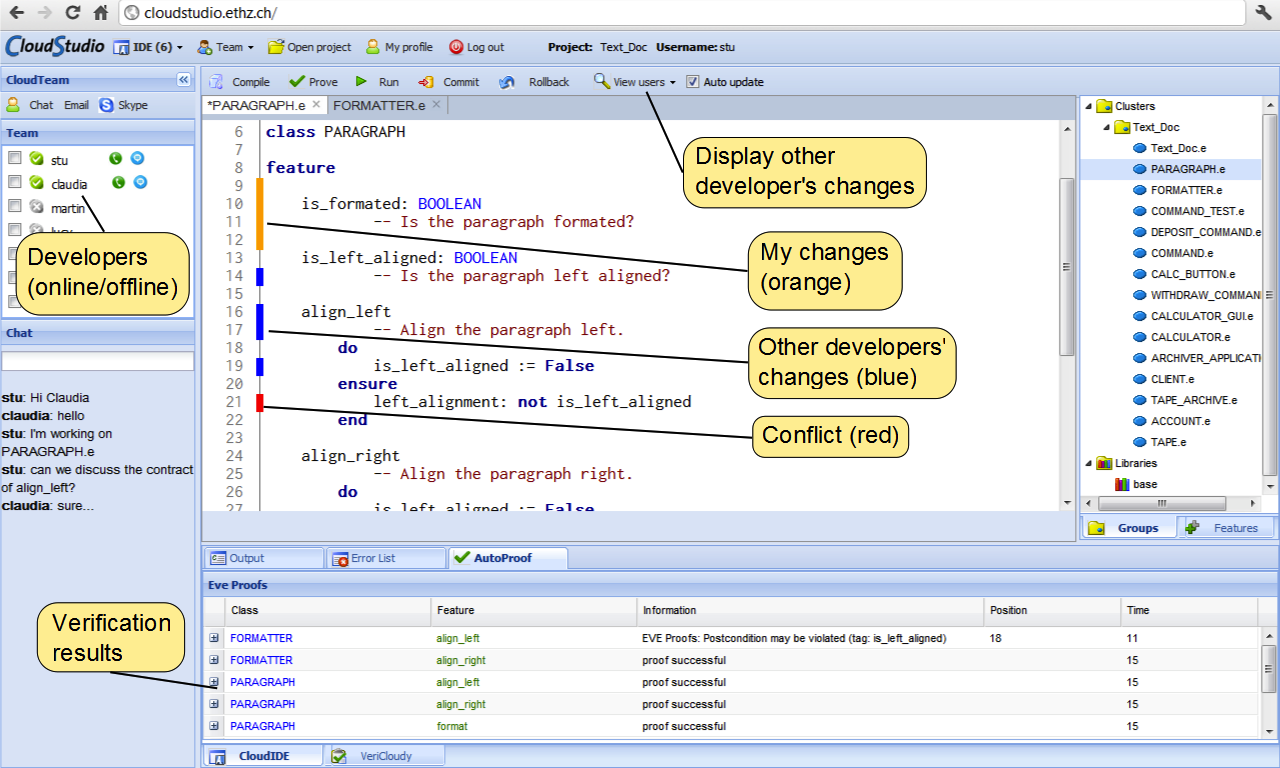} 
\end{center}
\caption{Main CloudStudio window for user Claudia (balloons are not part of the user interface but of this figure's caption).}%
\label{fig:CloudStudio}%
\end{figure}

This section gives an overview of CloudStudio from the perspective of two users---Stu and Claudia---who are working on the same project from different locations. 
Figure~\ref{fig:scenario} elaborates a usage scenario based on the fundamental
CloudStudio's features discussed in this section.

After logging in on \url{cloudstudio.ethz.ch} and selecting a project, Stu reaches the main CloudStudio window, pictured in Figure~\ref{fig:CloudStudio}.
The central frame shows the source code for the current class (\lstinline|PARAGRAPH|), which Stu can change with the class browser in the right-hand vertical frame.
The bottom frame displays the results of the latest compilation and verification runs.

Stu is editing class \lstinline|PARAGRAPH| concurrently with Claudia, who is working at a different location.
At any time, Stu can show or hide Claudia's changes to the code by toggling a button.
When changes are shown, as they are in Figure~\ref{fig:CloudStudio}, vertical bars of different colors mark each line of code according to its edit status: orange for lines changed or added by Stu (the current user); blue for lines changed or added by Claudia; red for lines with conflicts, that is edited differently by Claudia and Stu; lines without a colored bar are unchanged by anyone.
When he compiles the project, Stu can target the base version of the code (only unchanged lines), or include his or Claudia's changes to it, or both.
This mechanisms make Claudia and Stu aware of each other's work; they do not
have to block and immediately resolve conflicts, but they can continue working
without stomping on each other's feet. 

CloudStudio offers tools not only to detect and prevent conflicts, but also to resolve them.
Stu can see that Claudia is online in the left-hand top frame; he can call her on Skype, or chat with her directly in the left-hand bottom frame.
After agreeing on what to do with the conflicts, Stu clicks the \emph{commit} button to force a synchronization with Claudia.
CloudStudio's explicit commit works quite differently than in standard IDEs: the advanced features for configuration management make its usage quite infrequent.
When Stu commits, CloudStudio synchronizes the base version to Stu's current version; if lines with conflicts remain, the commit conservatively skips them so that the base version is in a consistent state.
The IDE shows the current base version number $n$---corresponding to ``unchanged'' lines of code in the editor---in the top-left corner (\textbf{IDE ($n$)}).

Chatting and talking can become more effective if Claudia and Stu have a means to type collaboratively on the very same piece of code, and to see it change to reflect the edits by both.
To this end, CloudStudio offers the \emph{interweave} editing mode where participants work on the code as if they were sitting at the same keyboard.
With interweave mode on, the notion of conflict disappears, because Claudia and Stu are effectively working in the same editor, similarly as in GoogleDocs.
Stu can enable or disable interweave editing at any time.
In fact, most of the development is carried out without interweaving, which is appropriate for fine-grained conflict resolution but generates too much jumble if used for most concurrent editing.

On top of the tools for collaborative development, CloudStudio features a
standard IDE integrated in the browser.
It even offers tools for automated
verification, so 
Stu can inspect failed verification attempts and accordingly modify the code to correct errors.

\begin{FPfigure}
  \label{fig:scenario}
  \includegraphics[width=\textwidth]{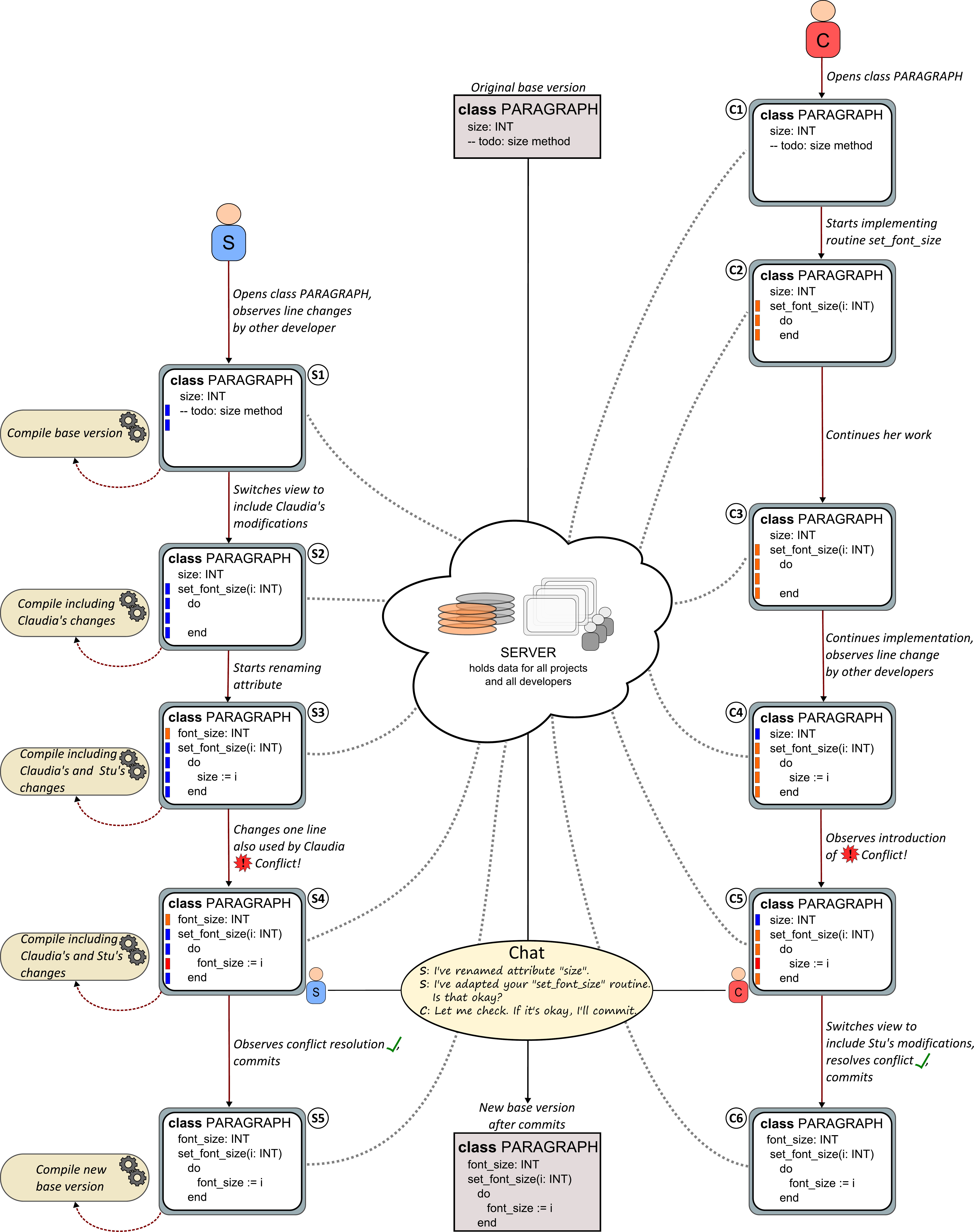}

  \caption{Programmer Claudia (rightmost column) starts working on a class
  \mbox{\lstinline|PARAGRAPH|.} CloudStudio displays the class current
  base version as plain text in Claudia's browser (C1); once she
  starts modifying the class, it marks in orange the code added that
  is not committed yet (C2--C3).  In the meanwhile, programmer Stu
  (leftmost column) also starts working on the same
  \lstinline|PARAGRAPH| class. Stu notices that the ``to do'' comment
  line is marked blue (S1); this means that Claudia has modified that
  line. Stu switches view to see exactly Claudia's work (the
  implementation of \mbox{\lstinline|set_font_size|}), also displayed
  marked in blue (S2). Fully aware of Claudia's concurrent editing,
  Stu does some light refactoring, consisting of renaming attribute
  \lstinline|size| to \lstinline|font_size| (S3). Claudia is aware of
  the change, because attribute \lstinline|size|'s line is marked blue
  in her editor (C4).  At this point, Stu tries to compile his current
  view of the project; compilation fails because of the different
  attribute names in his and Claudia's combined edits.  Stu easily
  figures out the problem and decides to fix it himself. This
  introduces a line marked red inside the routine Claudia has created,
  to denote a line modified differently by the two users with respect
  to the base version (S4).  CloudStudio makes reconciling the two
  versions very easy, in terms not only of programming but also of
  coordination between developers.  In fact, Stu concisely tells
  Claudia about the problem and how he solved it.  There is no need for
  complex communication, because both developers are aware of which
  parts have been changed by whom and how (S4--C5).  After quick
  agreement, Claudia and Stu decide to synchronize the base version
  for other developers (S5--C6).  Further modifications can now rely
  on a conflict-free up-to-date version of the class.}
\end{FPfigure}

\section{Unobtrusive Configuration Management and Awareness}\label{configuration}

This section presents the major feature offered by CloudStudio to support collaborative development: a configuration management system that is not centered around the rigid notion of revision, and that facilitates concurrent collaborative work by multiple programmers.

\subsection{Overview of the Problem}

The goal of configuration management is to track and control the evolution of
software artifacts---code, imprimis---during project development. The evolution
is three-dimensional, since software evolves in time, across developers, and in
different modules.

The standard approach to configuration management---implemented by tools such as
CVS and Subversion---uses a client/server architecture, where a central
repository stores incremental snapshots of the codebase, and every developer is
a client of the repository who maintains a local working copy of the code.
Synchronization between working copies and the central repository occurs by
explicit client request through \emph{update} and \emph{commit} operations. When
a client $A$ commits, the content of the central repository is changed to
include $A$'s changes present in its working copy. Conversely, when a client $B$
updates, $B$'s working copy is updated to coincide with the central
repository's. Even if so-called distributed version control systems---such as
Git---do not use a client/server architecture, they still require manual
operations, comparable to updates and commits, to synchronize a local copy with
others. This paradigm makes \emph{conflicts} likely to occur whenever two
developers work on the same portion of code without being aware of each other:
their local copies may diverge in irreconcilable ways, hence they have to
undergo a painful process of analysis and coordination to produce a unique consistent
version of that piece of code.

CloudStudio targets the shortcomings of traditional CM systems to facilitate
collaborative development by abandoning the update-commit paradigm and by
integrating an awareness system of what other developers are doing in the IDE.
This way, developers using CloudStudio never have to update, and commit only
very infrequently, while being constantly aware of potentially conflicting edits
as they set in, before fixing them becomes too burdensome.

\subsection{Configuration Management Model}
CloudStudio stores the current base version of a project's code in a relational database hosted ``in the cloud''.
The database table consists of the four attributes:
$$
\langle \textsc{File}, \textsc{Line\#}, \textsc{Text}, \textsc{Owner} \rangle\,,
$$
which respectively indicate a source file name, a line number in that file, the text appearing at that line number, and which users (if any) are editing that line.

The database stores the \emph{base} version of the codebase with a tuple:
$$
\langle f, k, l, \perp \rangle
$$ 
for\ \ each\ \ line\ \ $l$\ \ in\ \ position\ \ $k$ in\ \ a\ \ project file $f$, where $\perp$ denotes \emph{base} versions.
For example, if\ \ the\ \ fifth\ \ line\ \ of\ \ file\ \ ``stack.e''\ \ contains\ \ the\ \ signature\ \ \ of routine \mbox{\lstinline|push (v: INTEGER)|}, the database will store the tuple $\langle \text{stack.e}, 5, \text{\lstinline|push (v: INTEGER)|}, \perp \rangle$.
Whenever a user $u$ changes the line in position $k$ in a project file $f$ into the string $l'$, the database adds the tuple:
$
\langle f, k, l', u \rangle
$, which records $u$'s version of the line.

Since a tuple is added for every user who edits a line, we can search for \emph{conflicts} by looking up tuples that only differ in the last component, with two values other than $\perp$.
For example, if Claudia changes \mbox{\lstinline|push|'s} argument type to \lstinline|ANY|, and Stu makes \mbox{\lstinline|push|} return a \lstinline|BOOLEAN| to signal whether the operation was successful, there is a conflict signaled by the two tuples $\langle \text{stack.e}, 5, \text{\lstinline|push (v: ANY)|}, \text{Claudia} \rangle$ and\\ $\langle \text{stack.e}, 5, \text{\lstinline|push (v: INTEGER): BOOLEAN|}, \text{Stu} \rangle$.

Whenever a user $u$ performs an explicit \emph{commit}, the base version of the project is updated to reflect $u$'s latest edits.
That is, for every tuple $\tau = \langle f, k, l', u \rangle$ in the database \emph{without conflicts}, CloudStudio discards every tuple $\langle f, k, l, \perp \rangle$ (for every $l$), and replaces $\tau$ with $\langle f, k, l', \perp \rangle$.
If $\tau$ has conflicts, the base version of that line does not change.
Every commit generates a new base version of the project in the database; the
previous base version is purged from the database but it can be stored in a
back-end repository, allowing developers to roll back to older stable snapshots
of the project and to re-populate the database with them.

If two users $u_1, u_2$ are working in \emph{interweave} mode, CloudStudio stores their edits in the same tuples; that is, if either $u_1$ or $u_2$ changes the line in position $k$ in file $f$ into $l'$, the database stores the tuple $\langle f, k, l', \{u_1, u_2\}\rangle$.\footnote{The straightforward details of how this is implemented with relational schema are not discussed.}
Correspondingly, conflicts may arise between $u_1$ and $u_2$'s edits and somebody else's but not between $u_1$ and $u_2$.
Also, a commit by either one of $u_1$ and $u_2$ has the same effect of updating the base version to coincide with $u_1$ and $u_2$'s.

\subsection{Awareness System}
CloudStudio's awareness system extracts information from the configuration management database and displays it according to user preferences.
The basic behavior is that the editor shows the current user's edits, and the base version of every line untouched since the last explicit commit.
Each user retains ownership of her uncommitted changes; others can see them but not modify or commit them.

On top of this, CloudStudio provides options to see the changes introduced by other developers.
Each company and project has its own rules. The CloudStudio vision carefully refrains from imposing a specific methodology or process model, but provides the means to support such choices. 
The current user can select any other developer $u$ and choose to:
\begin{itemize}
\item Display all changes introduced by $u$;
\item	Display where $u$ introduced changes but do not show them;
\item	Display only where $u$'s changes generate conflicts;
\item	Do not display changes by $u$ at all;
\item Work in \emph{interweave} mode with $u$.
\end{itemize}
The last option avoids the introduction of conflicts and allows developers to modify lines collaboratively, in a way similar to GoogleDocs but with a level of granularity and control suitable for software development.

\section{Case Study} \label{case_study}

This section presents a case study that compares the performance of
two-programmer teams using CloudStudio against traditional CM
practices.  The overall goal of the case study is to assess the
usability of CloudStudio and its advantages for collaborative
development over traditional IDEs and CM techniques.

\subsection{Development Tasks}
The case study included three program development tasks, two focused on
refactoring and one on testing; all applications were written in Eiffel.
\begin{description}

\item[R1:] Task R1 targets an application implementing a card game
  (the card deck and the game logic); the complete
  application includes 210 lines of code over 4 classes.  Task R1
  requires refactoring of three classes, and development of new
  functionalities by extending the refactored classes; the task is
  collaborative because the new functionalities must work with the
  classes after refactoring. Refactoring included: method and field
  renaming; enforcement of Eiffel coding standards (e.g.,
  capitalization, comments); re-arrangement of methods in groups
  (marked by the \lstinline|feature| Eiffel keyword) according to
  their functionalities; code extraction into a new class.

\item[R2:] Task R2 targets an application modeling a coffee vending
  machine; users of the application have basic options to select
  coffee, can pay and receive change. The application includes 230
  lines of code over 3 classes.  Task R2 is similar to R1 except that
  it targets the coffee machine application: R2 requires refactoring
  and development of new functionalities by extending the refactored
  classes.

\item[T1:] Task T1 targets the same coffee machine application as task
  R2. It requires development of new functionalities (namely, the
  option to add milk to the coffee, and the dispatch of different cup
  sizes) and writing of test cases that achieve 100\% code coverage on
  the new code. Task T1 is also inherently collaborative as the
  development of new functionalities and of test cases occur
  concurrently, according to the concept of \textit{test-driven pair
    programming}~\cite{Goldman2010}.
  
\end{description}

\subsection{Subjects and Experimental Setup}
The subjects used in the study were six PhD students from our research group. 
All of them are experienced Eiffel programmers who frequently develop with
EiffelStudio and Subversion (SVN) as part of their PhD research; none of them
had used CloudStudio before the study, had taken part in its development, or has
much experience with collaborative development.

We randomly arranged the six subjects in three pairs: Team1, Team2, Team3. Team1
first performed task R1 with CloudStudio and then task T1 with EiffelStudio and
SVN. Team2 first performed task R1 with EiffelStudio and SVN and then task T1
with CloudStudio. Team3 first performed task R1 with CloudStudio and then task
R2 with EiffelStudio and SVN.

Each team performed its sessions according to the following protocol. The two
team members sat at the opposite corners of a large table with their laptops
connected to the network. Before beginning, the second author (henceforth ``the
experimenter'') gave a brief (5 min.) introduction to CloudStudio to both
programmers at the same time, where he showed them how to log-in and the basics
of the CM system without any reference to the development tasks. Then, he gave
them a sheet of paper with a description of the task they had to perform (the
second task was introduced only after completion of the first).
The two programmers received identical instruction sheets and had to coordinate
in order to split the work between them.

During the study nobody other than the experimenter and the two programmers was
in the room.  The programmers were only allowed to use instant messaging to
communicate; their position in the room and the experimenter ensured that no
other communication channel was available. The experimenter did not interfere
with the programmers other than to clarify possible unclear points in the task
description (but this was never necessary).

There was no time limit to complete the tasks: each session continued until the
current task was completed (the experimenter checked completeness \emph{a
posteriori} by manual inspection of the codebase).  After each session, the
experimenter recorded the total number of words exchanged via instant messaging and the overall time spent to complete the task.
An \emph{a posteriori} analysis of the communication logs, discussed in Section~\ref{sec:postmortem}, supports the hypothesis that these two measures (words and time) are reasonable proxies for the actual amount of communication between the two programmers that took place during the experiments.

\begin{figure}[!tb]
\begin{center}
  \includegraphics[width=.8\textwidth]{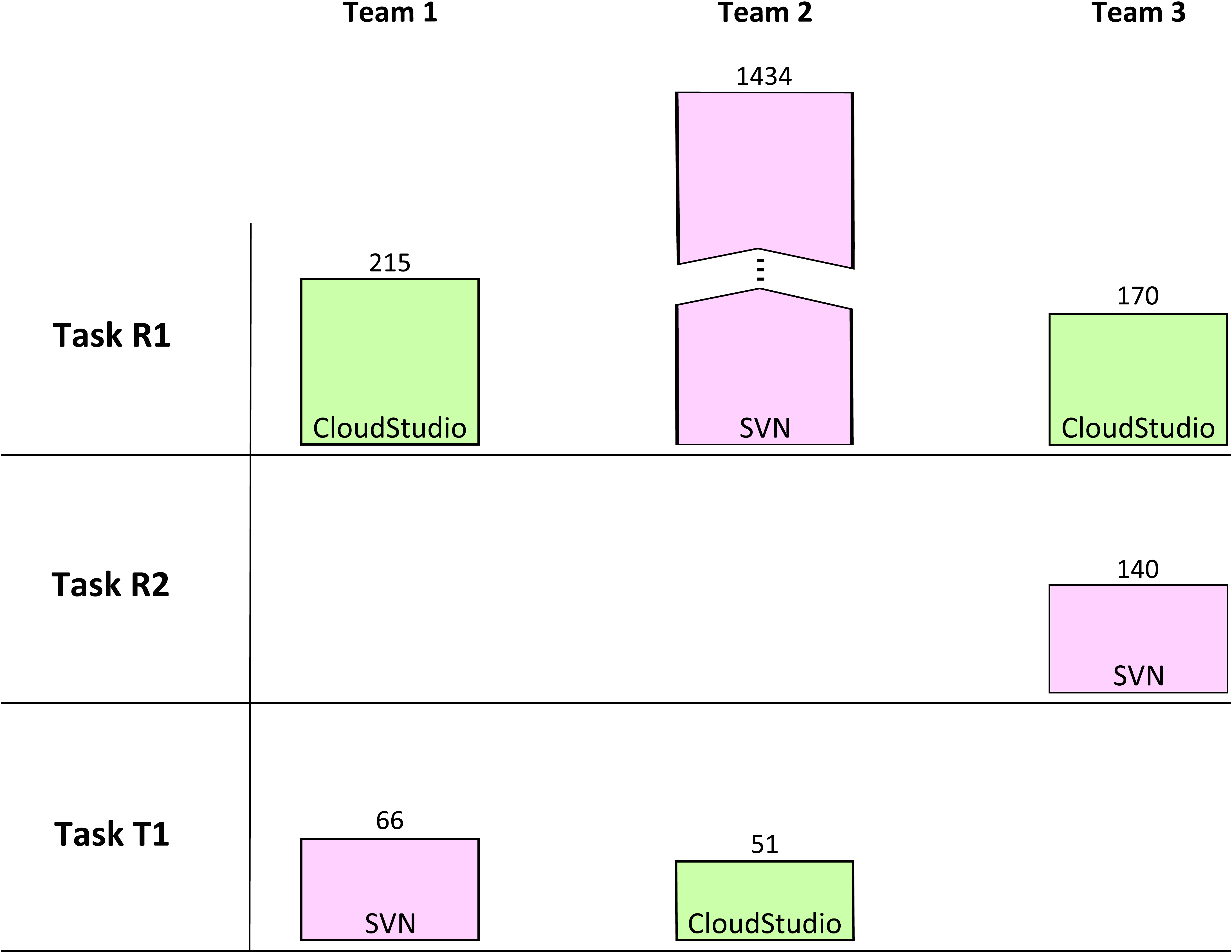}
  \caption{Results of the case study (the scale is not uniform).}
  \label{fig:comm_diagram}
\end{center}
\end{figure}

\subsection{Results}

Figure~\ref{fig:comm_diagram} reports the amount of communication between
programmers while performing the various tasks.  While all participants are
competent programmers, their speed and development style vary significantly; as
a result, the random assignment formed heterogeneous groups which may not be
directly comparable. The results in Figure~\ref{fig:comm_diagram}, however, show
a consistent advantage for teams using CloudStudio over teams using SVN: the
difference is sometimes small (as for task T1), sometimes conspicuous (as for
task R1 between Team2 and Team1); in all cases, CloudStudio required less
communication for the same task than SVN, even if the study's programmers used
it for the first time. Let us now describe the performance of the various teams
in more detail.

\textbf{Team1} delivered the best overall performance and was fluent both with
SVN and with CloudStudio; the two programmers worked well together and required
a limited amount of communication to synchronize properly. The comparison with
Team2 on the same tasks suggests that using CloudStudio is beneficial: Team1
outperformed Team2 almost by an order of magnitude when using CloudStudio on
task R1, whereas their performance became similar on task T1 where Team1 used
SVN. It was clear that Team1 was overall faster than Team2, but the
peculiarities of task R1 magnified the difference in favor of who could rely on
better collaboration tools.

The programmers in \textbf{Team2} had the greatest communication problems in the
study, as shown by their performance in task R1. The log of their message
exchanges shows that they had to debate several points of disagreement about how
to perform the refactorings, and that not being able to see in real-time what
the other was doing (as it happened when working with SVN) exacerbated their
disagreement and frustration.

Unlike the members of the other teams, the two programmers in \textbf{Team3}
worked with wildly different speed, to the point that in both tasks R1 and R2 a
programmer completed his part of the task when the other was still exploring the
system and understanding the instructions. The overall performance of Team3
required little communication in all cases, but this is mostly a result of the
fact that the different programmer speed forced a serialization between the two
programmers; hence, synchronization was not a big issue because the development
was not really collaborative and interactive.

We do not discuss in detail the \emph{time} taken by programmers because the
assignments emphasized correctness of the solution and did not pressure the
teams for time.  Anyway, and perhaps unsurprisingly, the overall time turned out
to be correlated with the amount of communication, hence all the experimental
data point to the same qualitative conclusions.

\subsection{Discussion} 
\label{sec:postmortem}

A \emph{post mortem} analysis of the instant messaging logs shows
recurring patterns of communications between programmers.  The initial
part of every session starts with a discussion of the task, after
which the two programmers negotiate a division of the labor and agree
on some synchronization mechanism.  During development with SVN,
messages such as ``Did you update your project?'' and ``I'm done with
implementing X and have committed'' are frequent.  With CloudStudio,
the same messages occurs much more sparingly, and some of the
remaining instances can probably be attributed to the programmers'
limited familiarity with CloudStudio and how it works (in fact, in
some cases of redundant notification messages using CloudStudio, the
recipient replied with sentences such as ``Just go ahead, I can see
your changes live'').

After the case study, we asked the participants to complete a simple
questionnaire about their experience and with requests for feedback.
The participants unanimously appreciated CloudStudio mechanisms for
the real-time visualization of other people's changes, and for the
immediate display of conflicts.  Disagreement existed on how severe a
problem are merge conflicts in everyday's software development: four
programmers consider it a serious hassle and appreciate better
mechanisms to prevent or manage conflicts; the other two maintained
that merge conflicts can be reduced to a minimum with a little
coordination.

In all, the participants to the study tend to agree with our conclusions that
CloudStudio offers valuable features for collaborative development and a more
flexible paradigm of CM. The generalizability of our results is necessarily
limited by the case study's scope and size, as well as by its reliance on
specific development tasks that emphasize real-time collaboration but may affect
only a limited part of large software projects. In this sense, the reaction of
one of the programmers in our study to task R1 is instructive: he was initially
skeptical and remarked that he ``would never do refactoring while another
programmer is implementing new functionalities''; after using CloudStudio,
however, he acknowledged that, with the right tools, such tasks can indeed be
performed in parallel.

\section{Other Features of CloudStudio's Prototype Implementation} 
\label{sec:verification}

A CloudStudio prototype is freely available at \url{cloudstudio.ethz.ch}; since it is entirely web-based, using it does not require downloading any software.
The implementation combines an editor written in Eiffel (automatically translated to JavaScript~\cite{E2JS}) with other functionalities implemented in Java using Google Web Toolkit v.~2.3, and leverages a MySQL database back-end.

CloudStudio currently supports development in Eiffel, but its
architecture is extensible to other programming languages such as C,
C\#, and Java.  Besides the innovative configuration management and
awareness system described in Section~\ref{configuration}, CloudStudio
offers the basic functionalities of traditional IDEs such as
EiffelStudio or Eclipse: an editor with syntax checking, a class
browser to navigate the project, and integration with the compiler.
At the time of writing, the complete implementation of interweave
editing is underway.

\begin{figure}[htp]
\begin{center}
  \includegraphics[width=\textwidth]{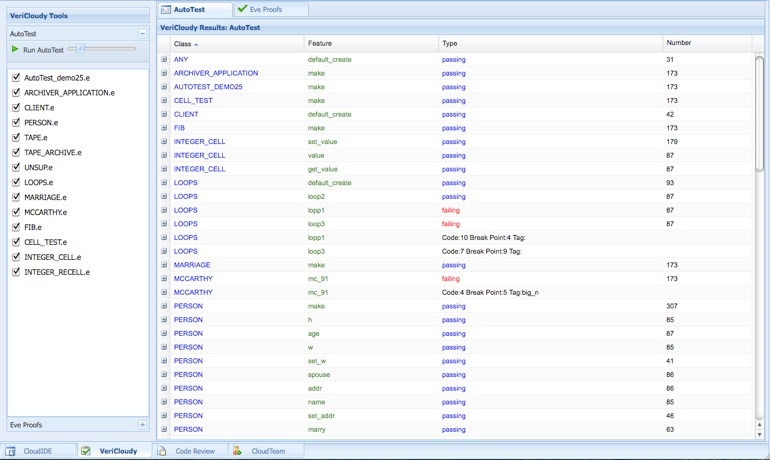}
  \caption{CloudStudio integrates the AutoTest framework for automatic
    random testing of object-oriented programs with contracts.
    AutoTest is completely automated: users only select the classes to
    be tested and the time allotted; AutoTest generates and executes a
    test suite for the classes. The figure shows an AutoTest report,
    which details passing and failing tests for every routine of every
    tested class. Clicking on an entry shows details about its
    associated test cases.}
  \label{fig:testing}
\end{center}
\end{figure}

\begin{figure}[htp]
\begin{center}
  \includegraphics[width=\textwidth]{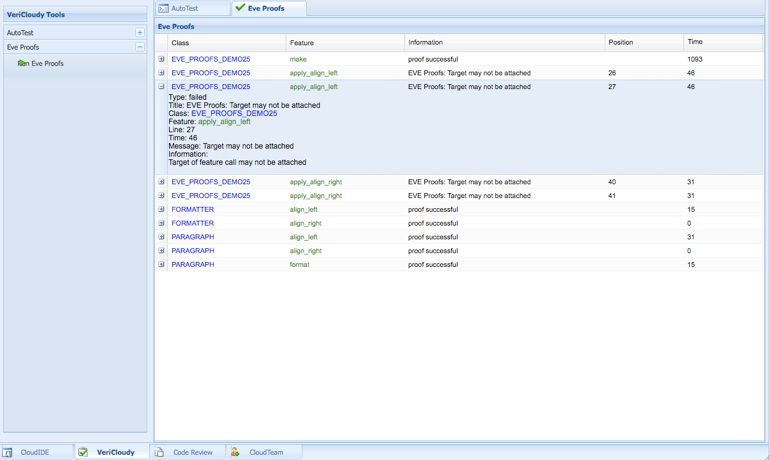}
  \caption[labelInTOC]{ CloudStudio integrates AutoProof, a fully
    automatic static verification tool that performs exhaustive
    correctness proofs of classes with contracts. AutoProof uses exact
    static analysis to establish if routines satisfy their
    postcondition for \emph{every} execution with input satisfying
    their precondition.  The figure displays an AutoProof report,
    showing, for every routine, whether its correctness proof
    succeeded.  AutoProof is sound but incomplete, hence failed proof
    attempts may be false positives that do not necessarily indicate
    the routine is incorrect.  A successful proof, on the contrary, is
    guarantee of correctness---within the limits of the given
    contracts.  }
  \label{fig:proofs}
\end{center}
\end{figure}

In continuity with our related work on formal verification centered
around EVE, the Eiffel Verification Environment
(\url{se.ethz.ch/research/eve/}), CloudStudio integrates verification
tools to help developers improve software quality.  It currently
supports testing with the AutoTest
framework~\cite{MeyerFivaCiupaETAL09} (see Figure~\ref{fig:testing}),
and formal correctness proofs with
AutoProof~\cite{NordioCalcagnoMeyerMuellerTschannen10} (see
Figure~\ref{fig:proofs}).  AutoTest performs random testing of
object-oriented programs with contracts, and it has proved extremely
effective in detecting hundreds of errors in production software;
AutoProof provides a static verification environment similar to
Spec\#\cite{SpecsharpCACM} but for Eiffel.  Both tools are fully
automatic and integrated with CloudStudio's CM system: testing and
proving sessions work on the current view selected by CloudStudio
users, which flexibly may or may not include concurrent edits by other
developers (as described in Section~\ref{configuration}).

\section{Related Work}
\label{sec:related_work}
The research community agrees on the potential impact that custom IDEs and collaboration tools can have on effective distributed software
development~\cite{Whitehead07}. 

Recent years have shown a trend towards supporting IDEs into the
web-browser. Some prominent examples of web IDEs are
Cloud9~(\bibentry{Cloud9IDE}), CodeRun~Studio
(\bibentry{CodeRunStudio}), and
Codeanywhere~(\bibentry{Codeanywhere}). As the browser is the natural
workbench for web applications, most web IDEs target languages for web
development (e.g., Javascript) rather than general purpose programming
languages such as Eiffel or Java.  Another limitation of most
commercial web IDEs currently available is their focus on supporting
run-of-the-mill functionalities that are standard in stand-alone IDEs,
as opposed to embracing new development and communication modes.

Some research prototypes of web IDEs have experimented novel approaches to collaborative development. Besides CloudStudio, the project Collabode~\cite{Goldman2011} supports real-time code sharing among developers through a web IDE.
Unlike CloudStudio, however, Collabode does not introduce new notions
of CM, and it is mainly intended for developers simultaneoutly working on the same piece of code with the same view (similarly to CloudStudio's interweave mode).

One of the most comprehensive frameworks for distributed
development is IBM's Jazz~\cite{Jazz}, built on top of the popular
Eclipse IDE. Jazz offers advanced communication and collaboration mechanisms,
but it is still built around the conventional CM model where files are the smallest unit of revision. Real-time collaboration (for example, for pair-programming) can be added on top of Jazz through tools such as Jazz Sangam~\cite{Vijay2008}, which, however, are separate entitites that do not fully integrate with the rest of the CM system.

A different small group of research tools such as Syde~\cite{HattoriLanza10} and
CollabVS~\cite{HegdeDewan08} introduce new models of CM, where more abstract and
flexible change analyses are possible. Syde works on abstract
syntax trees of the code, and defines changes as abstract operations on trees.
Crystal~\cite{BrunHolmesErnstNotkin11} is based on the idea of constantly trying
to merge the software artifacts of different developers in order to detect
conflicts as early as possible. These concepts are quite novel, yet still ultimately centered on the notion of conflict (and conflict resolution). CloudStudio's focus is instead on providing programmers with real-time code change awareness, which can prevent many conflicts from arising in the first place.

\section{Conclusions} \label{conclusions}

We have described a new paradigm of cloud-based software development,
addressing the needs of modern distributed projects, and presented the
first version of a supporting web-based tool called CloudStudio,
freely available for experimentation.


As can be expected with such a novel approach, many questions remain
open, providing both theoretical research challenges and practical
engineering goals. We plan to extend the new paradigm of configuration
management outlined above.
Much work remains on the IDE, in particular more
sophisticated display of the changes introduced by the
developers. More collaboration tools are needed, in particular to
support the new modes of code inspection made possible and desirable
by the Internet~\cite{Meyer09}.
%
%
We also intend to perform more extensive empirical evaluation of the
effectiveness of the ideas and tools, both in an industrial setting
and through systematic use of the tools in the multi-university
distributed \DOSE project and course on distributed and outsourcing
software engineering course~\cite{NordioMitinMeyer2010}.

\section*{Acknowledgments}

We thank Le Minh Duc, Alexandru Dima, and Alejandro Garcia for contributing to the prototype implementation of CloudStudio. 
Julian Tschannen and Yi (Jason) Wei contributed to the integration of the verification tools, and participated to the case study with Yu (Max) Pei, Marco Piccioni, Marco Trudel, Scott West.
CloudStudio's startup funding through the Gebert-Ruf Stiftung is gratefully acknowledged.

\bibliographystyle{abbrv}
\bibliography{literatur}

\end{document}